\documentclass[10pt,preprint]{aastex}

\begin{document}
\pagenumbering{arabic}

\title{THE FLATTENING OF GLOBULAR CLUSTERS}

\author{Sidney van den Bergh}
\affil{Dominion Astrophysical Observatory, Herzberg Institute of Astrophysics, National Research Council of Canada, 5071 West Saanich Road, Victoria, BC, V9E 2E7, Canada}
\email{sidney.vandenbergh@nrc.gc.ca}

\begin{abstract}

In the three nearest luminous galaxies, the Milky Way System,
the Andromeda Galaxy and NGC 5128 the brightest globular clusters are rounder than the faintest ones. On the other hand (contrary to some previous results) the flattening of individual LMC clusters is found to be independent of their luminosities. This suggests the possibility that the relationship between the flattening and luminosity of clusters might depend on host galaxy luminosity. No significant differences are found between the intrinsic flattening distributions of Galactic old halo, Galactic young halo and Galactic bulge/disk clusters. Such a dependence might perhaps have been expected if tidal forces (which are largest at small Galactocentric distances) removed angular momentum from globular clusters. The preliminary conclusion by Norris that clusters with blue horizontal branches are more flattened than red HB clusters is not confirmed by the larger data base that is now available. In other words there is no evidence for the puzzling claimed correlation between the flattening and the horizontal branch morphology of Galactic globular clusters.

\end{abstract}


\section{INTRODUCTION}

Many years ago it was noted (van den Bergh 1983a) that the
flattening of the clusters of all ages in the Large Magellanic Cloud is typically greater than that of Galactic globular clusters.
Furthermore Davoust \& Prugniel (1990) discovered that the most luminous globulars in M31 and the Galaxy are rounder than intrinsically fainter globular clusters. It is the purpose of the present note to re-investigate these conclusions by taking into account the fact that the apparent flattening of clusters can be strongly affected by patchy asymmetric foreground absorption (van den Bergh 1983b). Furthermore the database available to Davoust \& Prugniel will be expanded by including recent compilations of data on M31 globulars by Barmby et al. (2007), for Galactic globular clusters by Harris (1996) [which is updated at http://physwww.mcmaster.cas/~harris/mwgc.dat] and by Mackey \& van den Bergh (2005). The observations by Harris et al. (2006) also provide information on the flatteening of globular clusters in the nearby giant elliptical galaxy NGC 5128. Finally old and new data on the ellipticity of clusters in the Magellanic Clouds are combined to re-investigate  the dependence of cluster ellipticity on the luminosities and ages of clusters in the LMC and SMC. This study is facilitated by the fact that the metallicity (and hence then dust
content) of the Magellanic Clouds is low, so that the effects of asymmetric foreground absorption can be neglected.

A theoretical discussion of the interpretation of flattening
distributions of globular clusters, together with references to the literature on this subject, has been published by Akiyama (1991). He found that gravothermal contraction makes the inner regions of clusters rounder as they evolve. Furthermore, the outer regions of clusters are expected to become rounder with age due to the stripping of stars by external tidal fields. On the other hand, tidal fields might also be able to stretch clusters and make them more elongated. Finally Goodwin (1997) has pointed out that strong tidal fields might rapidly destroy velocity anisotropies in initially tri-axial rotating globular clusters. Mergers might produce highly flattened clusters. However, the absence of binary Galactic globular clusters, and the paucity of young binary clusters like {\it h} and $\chi$ Persei, suggests that this process may not have been an important factor in shaping Galactic star clusters. In this connection it is of interest to note that the clusters NGC 6388 and NGC 6441, which might be regarded as possible merger suspects because they are composed of stellar populations with slightly different ages (Piotto 2008), are observed to be almost circular in outline with axial ratios of 0.99 and 0.98, respectively.

\section{Flattening and globular cluster luminosity}

\subsection{Flattening of Galactic globular clusters} 

Following Hubble (1936) the flattening of globular clusters will be defined as $\epsilon$ = {\it (a-b)/a}, where {\it a} and {\it b} are the major and minor axes of the cluster. Mackey and van den Bergh list values of $\epsilon$ for 94 globular clusters. The flattening values of Galactic globular clusters were derived by White \& Shawl (1987) from images in blue light using the Palomar and SRC Sky Surveys. A weakness of this database is that the derived cluster flattening values do not all refer to a standard isophote such as the cluster half-light radius. Mackey and van den Bergh (2005) list values of $\epsilon$ for a total of 94 Galactic globular clusters. Two of these objects, $\omega$ Centauri = NGC 5139 and M54 = NGC 6715 are widely regarded as being the stripped cores of now defunct dwarf spheroidals and will therefore be omitted from the present study. Data on the ellipticity of all remaining Galactic globular clusters for which this information is available are plotted in Figure 1. This figure clearly shows that the faintest Galactic globular clusters are also the flattest ones. Furthermore the data in the figure strongly hint at the possibility that the most strongly reddened Galactic globular clusters (which are plotted in red) may appear more flattened than the less reddened Galactic globular clusters (plotted in blue).  In the the present analysis are all clusters with $A_{v}$ $>$ 1.0 mag [$A_{v}$ = 3.1 E(B-V) assumed] have been excluded because their apparent flattening might have been affected by patchy foreground absorption. The most blatant example of this effect is provided by M19 (= NGC 6273), which has $A_{v}$ = 1.27 mag and is the flattest ($\epsilon$ = 0.27) known Galactic globular cluster. It suffers heavy absorption along its eastern edge (van den Bergh 1982a). M19 is also observed to exhibit strong differential internal reddening (Harris, Racine \& deRoux 1976), but according to unpublished observations by Rosino, quoted by Coutts Clement \& Sawyer Hogg (1978), it shows little flattening at infrared wavelengths. A Kolmogorov-Smirnov test shows a 91\% probability that Galactic globular clusters with $A_{v}$ $>$ 1.0 mag appear, on average, more highly flattened than those with $A_{v}$ $<$ 1.0 mag. This result justifies a strong suspicion that the apparent flattenings of highly reddened clusters have been affected by asymmetric foreground absorption. It therefore seemed prudent to omit such highly absorbed clusters from our discussion of the intrinsic flattening distribution in all nearby galaxies.

Data on the flattening distributions of the 54 Galactic
globular clusters having both $A_{v}$ $<$ 1.0 mag, and published $\epsilon$ values, are collected in Table 1. The data in this table, which are plotted in Figure 2,  show that intrinsically faint Galactic globular clusters with $M_{v} >$ -7.0  are flatter than more luminous ones having $M_{v} <$ -7.0. A Kolmogorov-Smirnov test shows that there is only a 3\% probability that the luminous and the faint cluster samples were drawn from the same parent population. This conclusion strengthens and confirms a similar result by Davoust \& Prugniel (1990) who, however, neglected to take into account the effects of absorption on the apparent flattening of globular clusters.

\subsection{Flattening of clusters in M31}

The data in Table 1 show that the flattening of globular
clusters in M31 (Barmby et al. 2007) also depends on luminosity.
Adopting $(m-M)_{o}$ = -24.4 and $A_{v}$ = 3.1 E(B-V) one finds that (for objects with $A_{v}$ $<$ 1.0 mag) faint clusters with $M_{v}$ $>$ -8.0 are more flattened than are the more luminous ones with $M_{v}$ $<$ -8.0. A K-S test shows that there is only a 1.5\% probability that the faint and the luminous M31 clusters were drawn from the same parent distributions of flattening values. The M31 globular cluster system therefore resembles the one surrounding the Milky Way, in which the intrinsically most luminous clusters are also found to be the roundest.

\subsection{Flattening of clusters in NGC 5128}

 A similar result (see Table 1) is also found in the giant elliptical
galaxy NGC 5128. Using $\epsilon$ measurements from Harris et al. (2006) and $M_{v}$ values given in van den Bergh (2007) one finds that the clusters with Mv
$>$ -7.0 are flatter than those with  $M_{v}$ $<$ -8.0. [None of these clusters is heavily reddened, so that the observed flattening values will not be greatly affected by patchy foreground absorption.] A Kolmogorov-Smirnov test shows that there is only a 0.1\% chance that the flattening distributions of the bright and faint samples of clusters in NGC 5128 were drawn from the same parent distribution. It is particularly noteworthy that all 13 clusters in NGC 5128 with $\epsilon$ $>$ 0.20 are fainter than $M_{v}$ = -7.5.  An obvious caveat about the present discussion of the flattenings of globular clusters in NGC 5128 is that the globulars with  $M_{v}$ $>$ -7.0 are very faint (V $\sim$21), so that their individual measured $\epsilon$ values may be subject to quite large random errors. Taken at face value the present results for the Galaxy, M31 and NGC 5128 suggest that the faintest globular clusters associated with giant galaxies are intrinsically flatter than are the most luminous globulars associated with these objects. It would clearly be important to strengthen and confirm this preliminary conclusion by obtaining flattening data for the globular clusters hosted by other relatively nearby luminous galaxies.

\subsection{Flattening of star clusters in the Magellanic Clouds}

Data on the ellipticities of star clusters in the Large Magellanic
Cloud have been taken from Geisler \& Hodge (1980), Frenk \& Fall (1982), Geyer et al. (1983), Zepka \& Dottori (1987), Kontizas et al.(1989) and Bhatia \& MacGillvray (1989). The averages of the published flattening values for each LMC cluster are listed in Table 2. Data on the flattenings of 34 clusters in the Small Magellanic Cloud have been published by Kontizas et al.(1990). For the SMC the flattening given in Table 3 refers to its value at the cluster half-light radius. The photometry of the SMC clusters in Table 3 is from van den Bergh (1981). From the agreement between independent flattening observations of the same clusters in the LMC it is estimated that the mean error of dividual estimates of $\epsilon$ is $\sim$0.09, or $\sim$0.07 if the data by Bhatia \& MacGillvray are excluded. From the rather limited database on clusters with UBV photometry that have been observed independently by two observers one gains the impression that the observational errors of cluster flattening determinations are about twice as large for faint clusters with V $>$ 12 than they are for the bright clusters having V $<$ 12. The rather large observed scatter in the ellipticities assigned to individual clusters is presumably due to the following
factors: (1) The measured ellipticities of clusters frequently depend on distance from the cluster center. Some dispersion will therefore arise unless all measurements refer the same isophote {\it i.e.} that which encloses half of the cluster luminosity in projection. (2) Background subtraction may be a problem for the least luminous clusters in the densest regions of the Magellanic Clouds. (3) Stochastic effects will affect all attempts to determine the shapes of the isophotes of all clusters, particularly those that are faint or highly resolved. Since only a single series of observations exists for the flattenings of SMC clusters (Kontizas et al. 1990) it is not possible yet to derive an independent estimate for the errors in the quoted ellipticities of SMC clusters.

In the LMC cluster age determinations, on a scale from I (very
young) to VII (very old), were taken from Searle et al. (1980). These were supplemented by assignment to age class VII for all of the globular clusters (van den Bergh 2000, p.104) in the LMC. Contrary to a previous result by Fall \& Frenk (1984) the data, which are plotted in Figure 3, show no evidence for any correlation between the age class and the flattening of clusters in the LMC. Also given in Tables 2 and 3 are values of the reddening-free parameter Q = (U-B) - 0.72 (B-V) introduced by Johnson \& Morgan (1951). This parameter has good sensitivity to cluster age for young clusters, but may be affected by metallicity for the oldest clusters. In the tables uncertain values are followed by a colon. The data in in Table 2 and Table 3 are plotted in Figure 4. This figure shows no evidence for any correlation between LMC and SMC cluster ellipticity and the parameter Q, which may be regarded as a proxy for age. This  is so because young blue clusters have more negative Q values than do older ones.

Figure 5 shows a plot of the ellipticities of LMC clusters as a function of their luminosity. Contrary to a previous result by van den Bergh (1983a) the data that are now available show no evidence for a correlation between cluster luminosity and cluster flattening. This result is true for both globular clusters (shown in the figure as triangles) and for younger clusters, which are plotted as dots.

The data in Table 1 clearly show that the Galactic globular clusters
with $A_{v} <$ 1.0 mag are, on average, much less flattened than are those of all of the clusters in the LMC. A K-S test shows that the probability that the Galactic and LMC cluster flattening distributions were drawn from the same parent distribution is $<$ 0.01\%. Even the faintest little-reddened Galactic globulars are less flattened than the clusters in the LMC. A K-S test shows that there is only a 6\% probability that the 12 Galactic clusters with $M_{v} >$ -7.0 were drawn from the same flattening distribution as the LMC clusters. A comparison between all Galactic globular clusters with $A_{v} <$ 1.0 mag and the 10 objects in Table 2 which are classified as being either globular clusters (van den Bergh 2000, or that belong to Searle et al. (1980) age class VII, yields a probability of only 0.2\% that the LMC and Galactic globulars were drawn from the same parent population of flattening values. It is concluded that both globular clusters and younger clusters in the LMC are significantly more flattened than are Galactic globular clusters. Unfortunately little or no information is available on the flattening distribution of Galactic open clusters. However, casual inspection of the prints of the Palomar Sky Survey suggests that Galactic open cluster are mostly almost circular in outline. This suggests that Galactic open clusters resemble Galactic globulars and therefore differ systematically from their counterparts in the Clouds of Magellan. The reason for this systematic difference between Galactic star clusters and those in the Magellanic Clouds remains a mystery.

From the data in Table 1 it is seen that the clusters in the Small Magellanic Cloud are typically much more flattened than those of the globular clusters surrounding the Galaxy. A K-S test shows only $<$0.01\% probability that the flattening distributions of Galactic and Small Cloud clusters were drawn from the same parent population. Kontezas et al. (1990) found the star clusters in the SMC to be even flatter than those in the LMC. This conclusion is consistent with the data in Table 2 (LMC) and Table 3 (SMC). However, a Kolmogorov-Smirnov test shows that the observed difference in  flattening distributions does not reach a respectable level of  statistical significance. It is, however, of interesting to note that NGC 121, which is the single globular cluster in the SMC, has $\epsilon$ = 0.30 (Geyer et al. 1983), or $\epsilon$ = 0.28 (Kontizas et al. 1990) - which is  flatter than any of the globular clusters in the LMC - which all have $\epsilon$ $<$ 0.2. On plates taken in good seeing (van den Bergh 1983a) the cluster Hubble VII, which is the brightest globular in the dwarf galaxy NGC 6822, is also seen to be quite flattened. Clearly it would be important to obtain additional measurements of the flattenings of clusters in more dwarf galaxies to see if they follow the trend set by the LMC, the SMC and NGC 6822. The globular clusters that are associated with the Fornax and Sagittarius systems do not appear particularly flattened. It would be interesting to observe the cluster systems surrounding additional dwarf galaxies to see if the clusters that are associated with dwarf irregulars are systematically more flattened than those hosted by dwarf spheroidal galaxies. It would also be of interest to find out if other dwarf irregulars resemble the Magellanic Clouds in which both the young open clusters, and the old globular, are more flattened than similar objects associated with giant galaxies. The data in Table 3 are too scanty to determine if there is a correlation between the luminosity and the flattening of clusters in the SMC. However,it is of interest to note that the four brightest (V $<$ 11.3) Small Cloud clusters are all very flattened having $\langle$ $\epsilon$ $\rangle$ =0.26. If real, this trend would run counter to that in giant galaxies in which it the faintest clusters that are the most flattened.

\section{Cluster flattening and population type.}

Mackey \& van den Bergh (2005) have used the morphology of globular
cluster horizontal branches to assign these objects to different Galactic populations such as old halo (OH), young halo (YH) and bulge/disk(BD). Among clusters with A$_{v} <$ 1.0 mag (for which flattening is least likely to be affected by patchy foreground absorption) no statistically significant differences are found between the $\epsilon$ distributions of 31 old halo clusters, 13 young halo objects and 16 bulge/disk clusters.   It is noted in passing that the relaxation times of Galactic globular clusters (Webbink 1985) are uncorrelated with their flattenings, even though faint clusters are more flattened than luminous ones. The reason for this is that faint Galactic globular clusters are, on average, larger than luminous ones. The longer relaxation time of clusters containing a large number of stars is therefore approximately compensated for by the fact that that luminous clusters tend to be  more compact than dim ones. If tidal effects contribute to the flattening of globular clusters, then one might expect clusters close to the Galactic center to be more flattened than those in the outer Galactic halo. This is indeed observed to be the case. However, a K-S test shows that the difference in flattening distributions of clusters with $A_{v} <$ 1.0 having $R_{gc} <$ 12 kpc does not differ at a respectable level of statistical significance from that of the clusters with  $A_{v} <$ 1.0 mag and $R_{gc} >$ 12 kpc. It is noted in passing that the elimination of highly reddened globular clusters from the present sample has introduced a bias against clusters with collapsed cores which are strongly concentrated behind the dust clouds that shroud the Galactic center.

\section{Flattening and horizontal branch morphology}

It has been known for many years (van den Bergh 1965, 1967, and
Sandage \& Wildey 1967) that globular clusters exhibit a ``second parameter'' effect with clusters of similar metallicity showing differing population gradients along their horizontal branches. This effect has variously been attributed to differences in helium abundance, CNO group abundances or stellar rotation. Surprisingly Norris (1983) found an apparent correlation between cluster flattening and horizontal branch gradient, in the sense that (among clusters of intermediate metallicity) objects with blue horizontal branches can have any value of $\epsilon$, whereas nearly round clusters all have red horizontal branches. A better way to look into this problem is provided by using the larger and more recent sample of $\epsilon$ and HB-index values listed by Mackey \& van den Bergh. Their HB index is defined as (B-R)/B+V+R) in which B is the number of stars that lie to the blue of HB instability strip, V is the number of stars in this strip, and R is the number of stars to the red of the horizontal branch instability strip. After omitting (1) highly reddened clusters with $A_{v} >$ 1.0 mag (for which the apparent flattening might be due to asymmetric reddening),(2) intrinsically faint clusters $M_{v} >$ -6.0 (in which flattening measurements are intrinsically uncertain because of low total stellar content), and (3) $\omega$ Cen and M54 (which might be stripped galaxy cores), one obtains a sample of 54 Galactic globular clusters. Half of these clusters have a horizontal branch index $<$ +0.6 and half of them have an HB-index $>$ +0.6. A Kolmogorov-Smirnov test shows that there is no statistically significant difference between the ellipticity distributions of the Galactic globular clusters with red and with blue horizontal branches. Since the present sample is exactly twice as large as that used by Norris it is concluded that his result, which was significant at the 96\% level, was probably due to the well-known perversity of small-number statistics. In his original paper Norris considered only those globular clusters with intermediate metallicities in the range 1.4 $\leqslant$ [Fe/H] $\leqslant$ -1.9. If one applies the same restriction to the sample discussed above then one is left with only 29 clusters. For these objects a K-S test again shows no significant difference between the flattening distributions of the clusters with HB-index $<$ +0.6 and HB-index $>$ +0.6. It is therefore concluded that the best presently available data provide no evidence to support the conclusion by Norris (1983) that the flattening of Galactic globular clusters is correlated with their horizontal branch morphology.
   
About a quarter of all globular clusters exhibit an unusually
extended horizontal branch (Lee et al. 2007). This ``blue hook''
morphology probably indicates that such clusters had an unusual evolutionary history. On average these clusters are of above-average luminosity. A comparison between the distribution of the small population of little reddened blue hook clusters with a similar population of luminous globulars with normal horizontal branches shows no statistically significant difference in the distribution of cluster flattenings. It should, however, be emphasized that this conclusion is based on small samples.

\section{Conclusions}

Data in the larger database that is now available do not confirm Norris's (1983) surprising conclusion that the flattening of Galactic globular clusters correlates with their horizontal branch morphology.  Furthermore it is found that there is no difference between the flattening distributions among old halo, young halo and bulge/disk clusters (as defined by Mackey \& van den Bergh (2005). Finally the present data strengthen and confirm the conclusion of Davoust \& Prugniel (1990) that luminous Galactic globular clusters are, on average, rounder than are less luminous globular clusters. In this respect the Galaxy appears to resembles M31 and NGC 5128, but differs from the Magellanic Clouds (Frenk \& Fall 1982, van den Bergh 1983a, Goodewin 1997). The reasons for these difference are presently not understood. The conclusions listed above could be greatly strengthened by obtaining {\it (a-b)/a} values for Galactic globular clusters with $A_{v} >$ 1.0 at infrared wavelengths. Such flattening determinations would be much less affected by patchy foreground absorption than are existing measurements at shorter wavelengths. However, such measurements will not provide a panacea because the images of clusters in the infra-red are strongly affected by small numbers of cool red stars, whereas the images of clusters in blue light provide more nearly comparable contributions from red giants and blue horizontal branch stars. It would also be of interest to obtain flattening observations of globular clusters in other nearby dwarf galaxies. Such observations might allow one to see if they exhibit the same dichotomy between late-type dwarfs (hosting flattened clusters) and dwarf spheroidals (containing rounder clusters) that is hinted at by the dwarfs in the Local Group. Finally it would be of interest to obtain ellipticity measurements for a representative sample of Galactic open clusters to test the impression that these objects are (like Galactic globular clusters) rounder than their counterparts in the Clouds of Magellan.

I thank Bill Harris, Jun Ma, Thomas Puzia, and John Norris for helpful exchanges of e-mail correspondence. Technical support was provided by Brenda Parrish and Jason Shrivell. I am also indebted to a particularly helpful referee.

\newpage

\begin{figure}
\plotone{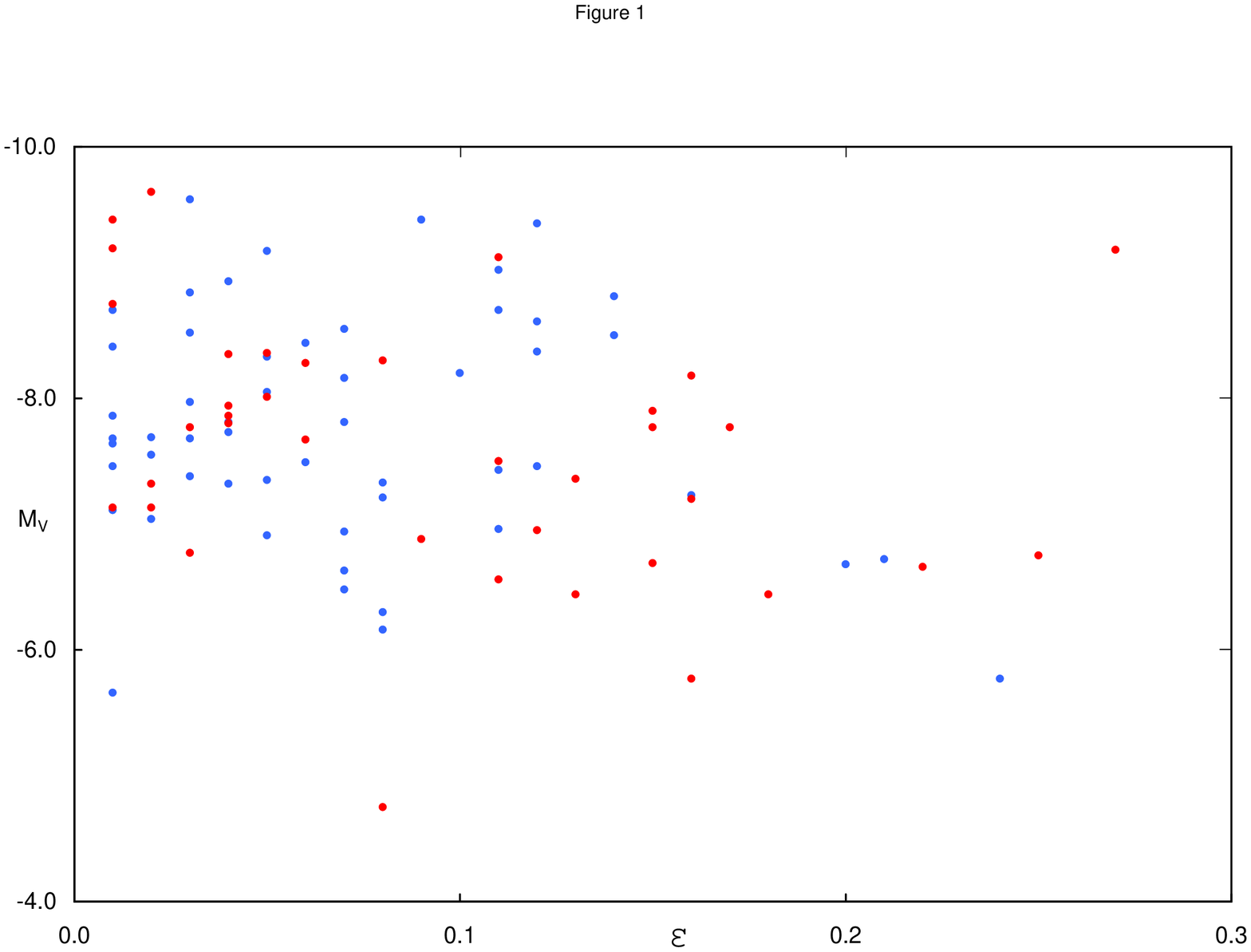}
\caption{Luminosity versus flattening for Galactic globular clusters. Clusters with $A_{v} <$ 1.0 mag are plotted in blue, and those with $A_{v} >$ 1.0 mag are drawn in red. The figure shows that the faintest Galactic globulars are also the most flattened. The figure also suggests that highly reddened clusters appear to be more flattened than less reddened ones. The red dot in the upper right hand corner of the figure represents the cluster M19.}

\end{figure}

\begin{figure}
\plotone{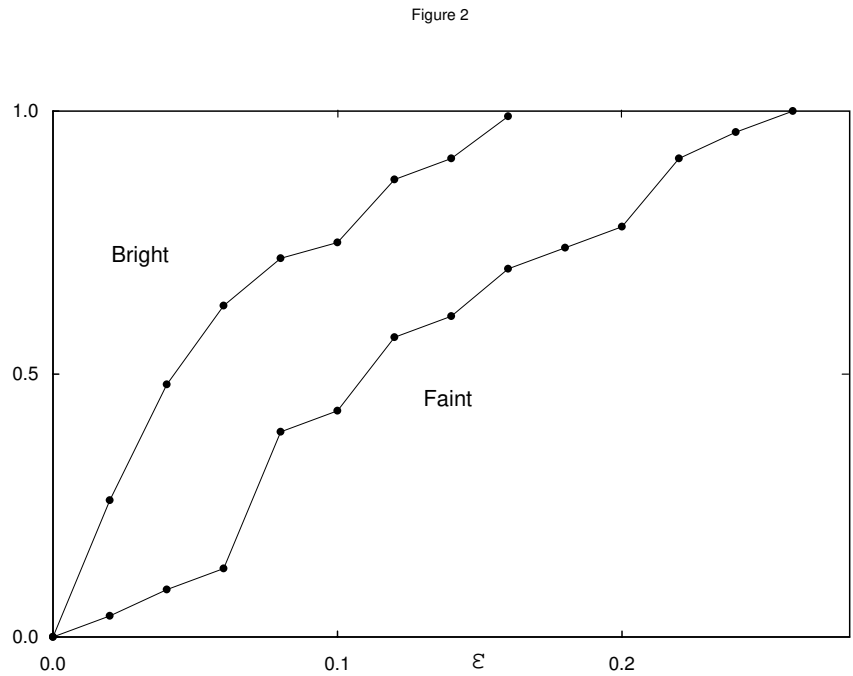}
\caption{Normalized frequency distributions of the apparent flattening values for luminous ($M_{v} <$ -7.0) and faint ($M_{v} >$ -7.0) Galactic globular clusters. [For reasons outlined in the text $\omega$ Centauri and M54 have been excluded from the samples plotted in Figures 1 and 2.] The figure shows that the apparent flattening values of the faint globular clusters are significantly greater than those of the more luminous ones.}

\end{figure}

\begin{figure}
\plotone{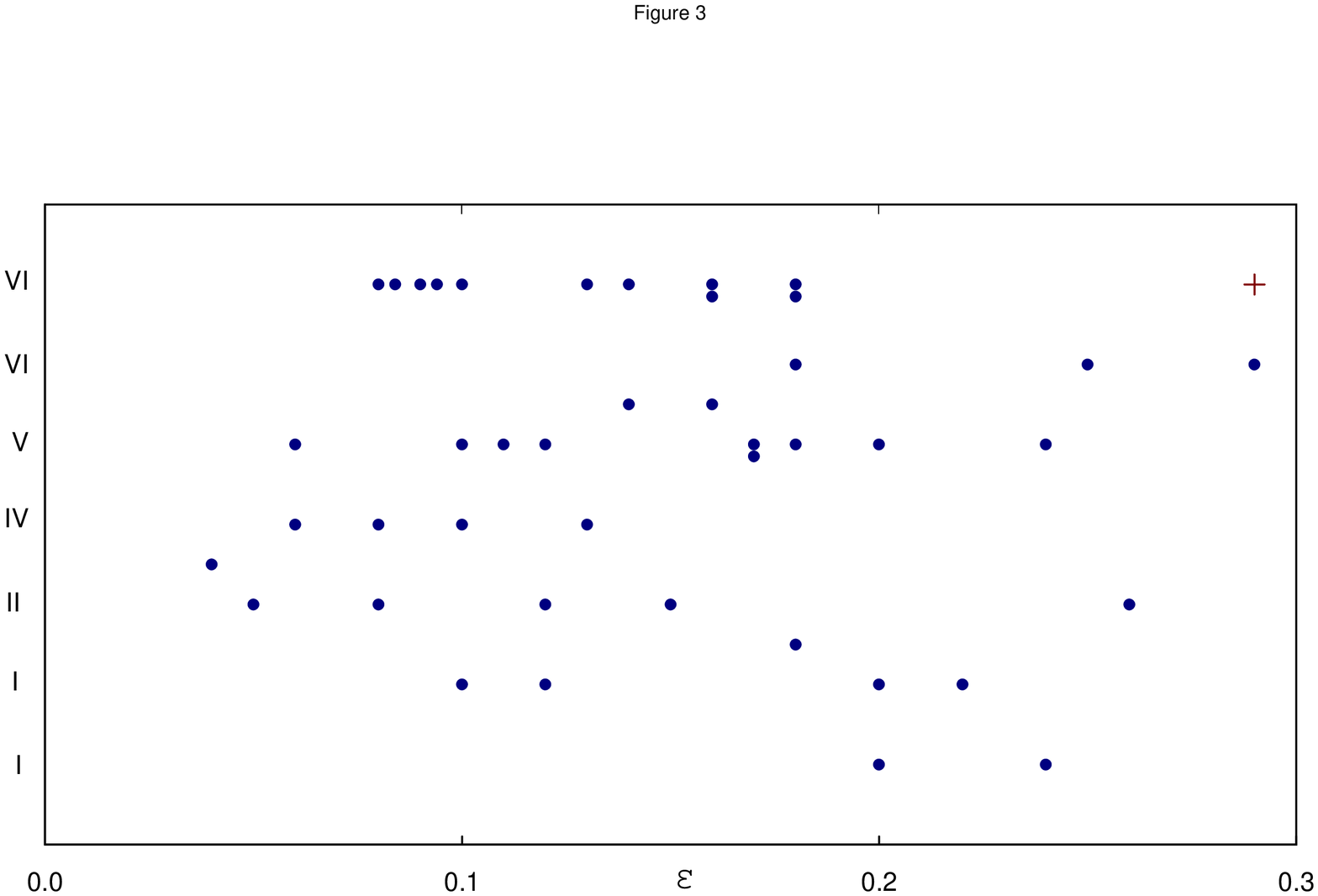}
\caption{Relation between the flattening and the Searle age class of globular clusters in the Magellanic Clouds. NGC 121 in the SMC is shown as a plus sign. The  figure shows no evidence for a dependence of cluster flattening on age, i.e. both the old and the young star clusters in the Magellanic Clouds are flatter than their Galactic counterparts.}

\end{figure}

\begin{figure}
\plotone{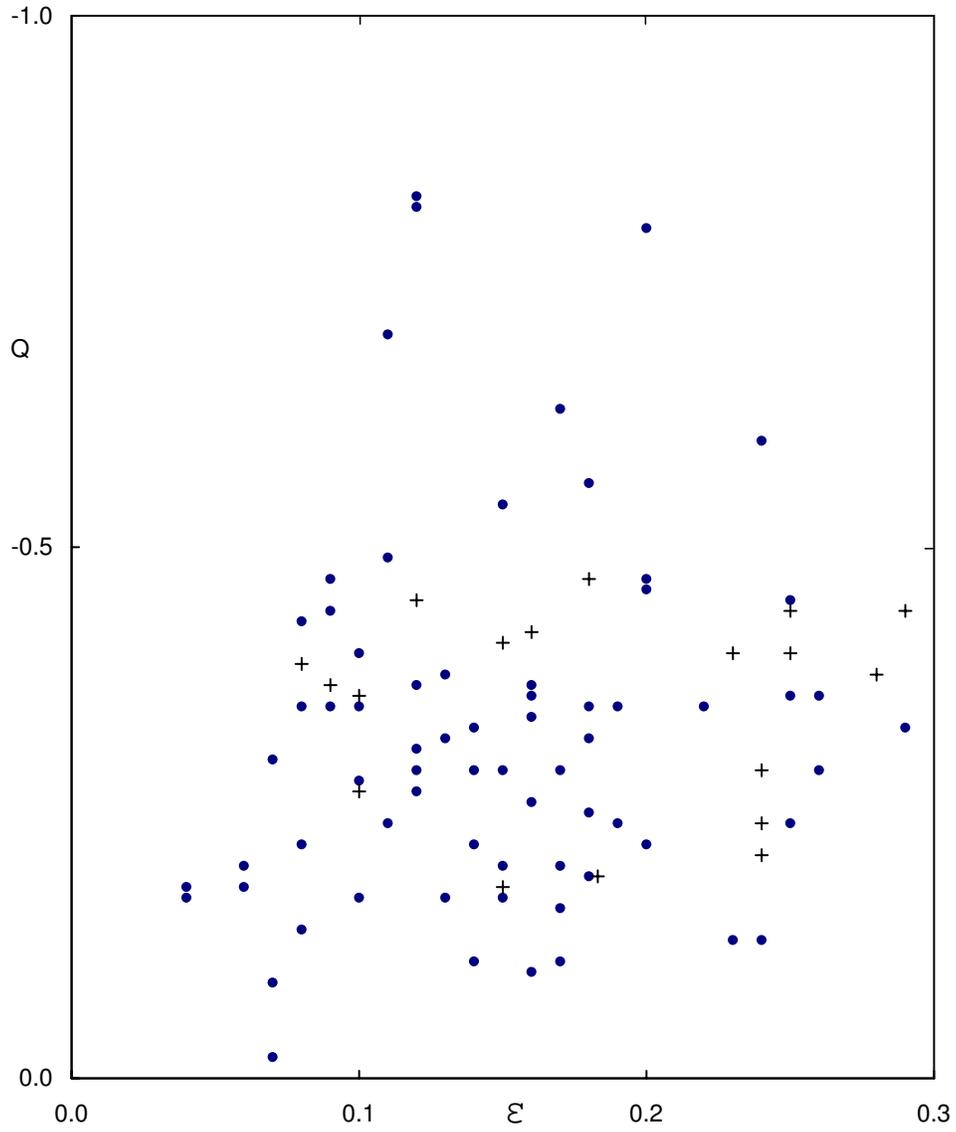}
\caption{Relation between ellipticity and the reddening-free parameter Q for clusters in the LMC (dots) and SMC (plus signs). In neither of these galaxies does the cluster flattening appear to be correlated with cluster age.}

\end{figure}

\begin{figure}
\plotone{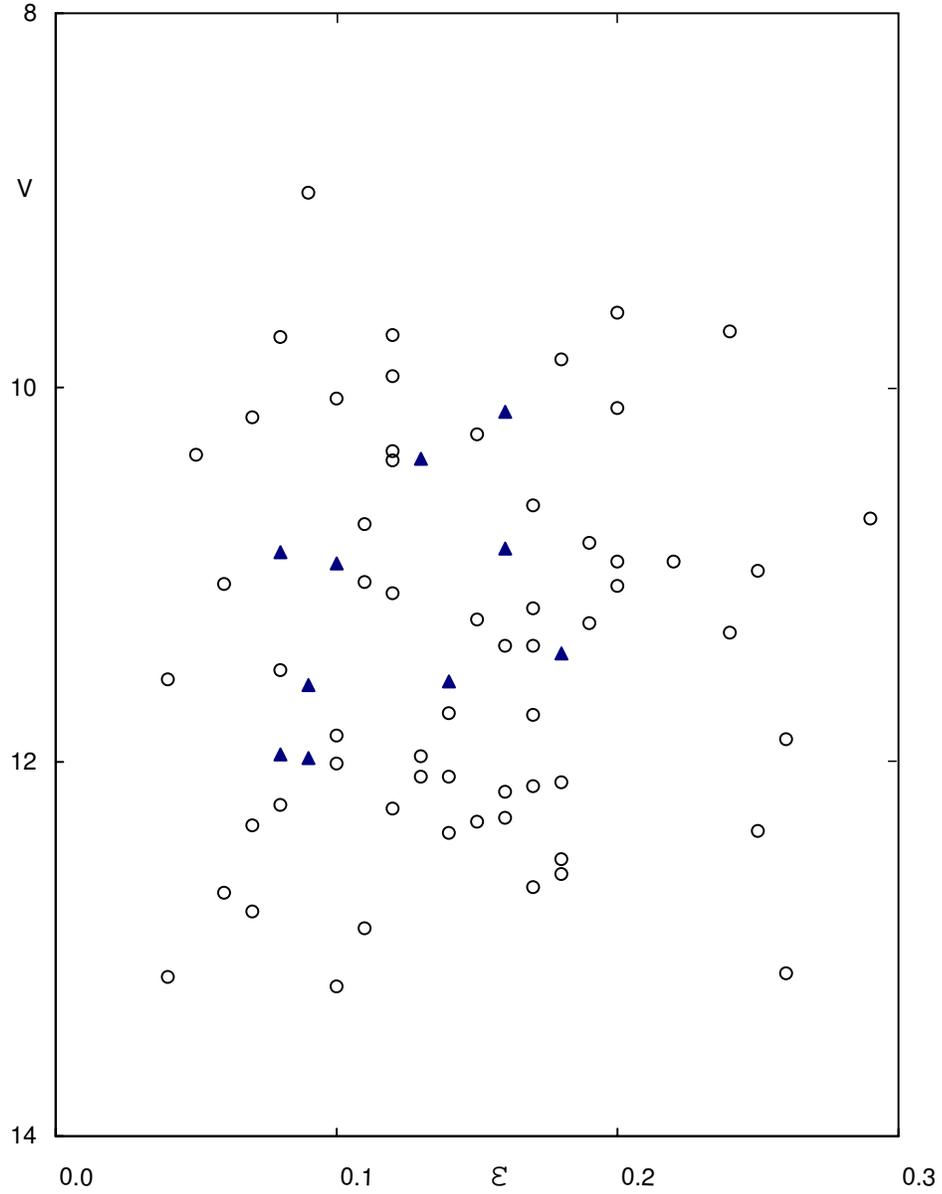}
\caption{Relation between luminosity and flattening of LMC clusters.
Globular clusters of Searle age class VII clusters are shown as triangles, younger clusters as circles. Contrary to some previous results the figure shows no evidence for a correlation between the luminosity and the flattening of clusters in the Large Magellanic Cloud.}

\end{figure}

\newpage

\begin{deluxetable}{lcccc}
\tablewidth{0pt}
\tablecaption{Normalized integral frequency distributions of flattening distributions for little-reddened globular clusters with $A_{v} <$ 1.0 mag}

\tablehead{\colhead{{\it (a-b)/a}} & \colhead{Galaxy}  & \colhead{Galaxy} &\colhead{M31} & \colhead{M31}}

\startdata
&luminous clusters &faint clusters &luminous clusters &faint clusters\\ &M$_{v}<$-7.0 &M$_{v}>$-7.0 &M$_{v}<$-8.0 &M$_{v}>$-8.0\\ \hline
$<$0.015  &    0.19     &         0.08   &       0.00     &      0.19\\
$<$0.035   &   0.38     &         0.08   &       0.00     &      0.19\\
$<$0.055   &   0.60     &         0.17   &       0.14     &      0.28\\
$<$0.075   &   0.71     &         0.42   &       0.24     &      0.33\\
$<$0.095   &   0.79     &         0.58   &       0.62     &      0.37\\
$<$0.115   &   0.86     &         0.67   &       0.76     &      0.42\\
$<$0.135   &   0.95     &         0.67   &       0.84     &      0.49\\
$<$0.155   &   0.98     &         0.67   &       0.84     &      0.63\\
$<$0.175   &   1.00     &         0.67   &       0.86     &      0.72\\
$<$0.195   &   1.00     &         0.67   &       0.89     &      0.79\\
$<$0.215   &   1.00     &         0.83   &       0.97     &      0.81\\
$<$0.235   &   1.00     &         0.92   &       0.97     &      0.88\\
$<$0.255   &   1.00     &         1.00   &       0.97     &      0.91\\
Total    &   n$=$42   &         n$=$12  &      n$=$37   &     n$=$43\\
\\
\hline
& NGC 5128   & NGC 5128    &LMC     & SMC\\
&luminous clsuters &faint clusters\\
     &Mv$<$-8.0          & Mv$>$-7.0\\
\hline
$<$0.015   &    0.04      &          0.03     &        0.00  &   0.00\\
$<$0.035   &    0.28      &          0.16     &        0.00  &   0.03\\
$<$0.055   &    0.49      &          0.29     &        0.03  &   0.03\\
$<$0.075   &    0.70      &          0.32     &        0.08  &   0.03\\
$<$0.095   &    0.81      &          0.42     &        0.17  &   0.09\\
$<$0.115   &    0.82      &          0.45     &        0.27  &   0.15\\
$<$0.135   &    0.86      &          0.58     &        0.37  &   0.24\\
$<$0.155   &    0.89      &          0.61     &        0.46  &   0.42\\
$<$0.175   &    0.96      &          0.61     &        0.61  &   0.45\\
$<$0.195   &    1.00      &          0.71     &        0.73  &   0.58\\
$<$0.215   &    1.00      &          0.74     &        0.81  &   0.64\\
$<$0.235   &    1.00      &          0.77     &        0.84  &   0.73\\
$<$0.255   &    1.00      &          0.81     &        0.90  &   0.94\\

Total    &    n$=$57     &         n$=$31    &      n$=$98  &  n$=$33\\

\enddata
\end{deluxetable}


\begin{deluxetable}{lllll}
\tablewidth{0pt}
\tablecaption{Mean flattening of NGC clusters in the Large Magellanic Cloud and for a few additional Large Cloud clusters for which photometry is available}

\tablehead{\colhead{Name} & \colhead{Age class} & \colhead{V} & \colhead{Q} & \colhead{$<\epsilon>$}} \startdata
N1466  &   VII  &   11.59  & -0.35  & 0.09\\
N1644  &   V    &   12.89  & -0.24  & 0.11\\
N1651  &   ...  &   12.67  & -0.20  & 0.17:\\
N1652  &   ...  &   13.13  & -0.29: & 0.26\\
N1696  &   ...  &    ...   &  ...   & 0.21\\
N1698  &   ...  &    ...   & -0.54: & 0.15\\
N1711  &   II   &   10.11  & -0.46  & 0.20\\
N1718  &   ...  &   12.25  & -0.29  & 0.12\\
N1734  &   ...  &    ...   &  ...   & 0.25\\
N1749  &   ...  &    ...   &  ...   & 0.32\\
N1751  &   V    &   12.11  & -0.19: & 0.18\\
N1754  &   GC   &   11.96  & -0.35: & 0.08\\
N1755  &   II-III  &  9.85 &  -0.32 &  0.18\\
N1783  &   V    &   10.93  & -0.22  & 0.20\\
N1786  &   GC   &   10.88  & -0.43  & 0.08\\
NN1786  &   GC   &   10.88  & -0.43  & 0.08\\
N1795  &   ...  &    ...   & -0.13: & 0.23\\
N1801  &   ...  &   12.16  & -0.10  & 0.16\\
N1805  &   ...  &   10.63  & -0.63  & 0.17\\
N1806  &   V    &   11.10  & -0.27  & 0.12\\
N1818  &   I    &    9.70  & -0.60  & 0.24\\
N1828  &   ...  &   12.52  & -0.25  & 0.18\\
N1831  &   V    &   11.18  & -0.11  & 0.17\\
N1835  &   VII  &   10.13  & -0.37  & 0.16\\
N1838  &   ...  &    ...   &  ...   & 0.17\\
N1839  &   ...  &   ...    & ...    &0.10\\
N1842  &   ...  &   ...    & ...    & 0.16\\
N1844  &   ...   &  12.08  & -0.29  & 0.14\\
N1846  &   V     &  11.31  & -0.13  & 0.24\\
N1847  &   ...   &  11.06  & -0.47  & 0.20\\
N1849  &   ...   &  12.80  & -0.02  & 0.07\\
N1850  &   ...   &   8.96  & -0.44  & 0.09\\
N1852  &   ...   &  12.01  & -0.28  & 0.10\\
N1854  &   II  &    10.39  & -0.37  & 0.12\\
N1856  &   IV  &    10.06  & -0.17  & 0.10\\
N1860  &   ... &    11.04  & -0.49  & 0.11\\
N1861  &   ... &     ...   &  ...   & 0.14\\
N1863  &   ... &    10.98  & -0.45  & 0.25\\
N1864  &   ... &     ...   &  ...   & 0.16\\
N1865  &   ... &     ...   &  ...   & 0.18\\
N1866  &   III &     9.73  & -0.22  & 0.08\\
N1868  &   ... &    11.56  & -0.17  & 0.04\\
N1870  &   ... &    11.26  & -0.35  & 0.19\\
N1871  &   ... &     ...   &  ...   & 0.20\\
N1878  &   ... &     ...   &  ...   & 0.21\\
N1885  &   ... &    11.97  & -0.32  & 0.13\\
N1897  &   ... &     ...   &  ...   & 0.12\\
N1898  &   GC  &    11.42  & -0.56  & 0.18\\
N1903  &   II  &    11.86  & -0.35  & 0.10\\
N1905  &   ... &     ...   &  ...   & 0.19\\
N1916  &   GC  &    10.38  & -0.38  & 0.13\\
N1917  &   ... &    10.25  & -0.17: & 0.15\\
N1943  &   III &    11.88  & -0.36  & 0.26\\
N1953  &   ... &    11.74  & -0.11  & 0.14\\
N1978  &   VI  &    10.70  & -0.33  & 0.29\\
N1983  &   ... &     9.94  & -0.83  & 0.12\\
N1984  &   ... &     9.72  & -0.82  & 0.12\\
N1987  &   IV  &    12.08  & -0.17  & 0.13\\
N2004  &   I   &     9.60  & -0.80  & 0.20\\
N2005  &   GC   &   11.57  & -0.33  & 0.14\\
N2019  &   VII  &   10.86  & -0.36  & 0.16\\
N2031  &   ...  &   10.83  & -0.24  & 0.19\\
N2038  &   ...  &    ...   &  ...   & 0.16\\
N2041  &   III  &   10.36  & =3D0.33  & 0.05\\
N2056  &   ...  &   12.34  & -0.09  & 0.07\\
N2065  &   III   &  11.24   &-0.29   &0.15\\
N2098  &   ...  &   10.73  & -0.70  & 0.11:\\
N2107  &   IV   &   11.51  & -0.14  & 0.08\\
N2108  &   ...  &   12.32  & -0.20  & 0.15\\
N2109\tablenotemark{*} &   VII  &    ...   &  ...   & 0.18\\
N2114  &   ...  &    ...   &  ...    0.42\\
N2116  &   ...  &    ...   &  ...   & 0.29\\
N2117  &   ...  &    ...   &  ...   & 0.41\\
N2121  &   VI   &   12.37  & -0.36  & 0.25\\
N2134  &   IV   &   11.05  & -0.20  & 0.06\\
N2135  &   ...  &    ...   &  ...   & 0.48\\
N2140  &   ...  &    ...   &  ...   & 0.27\\
N2154  &   V    &   12.13  & =3D0.19  & 0.17\\
N2155  &   VI   &   12.60  & -0.35  & 0.18\\
N2156  &   ...  &   11.38  & -0.16  & 0.17\\
N2157  &   ...  &   10.16  & -0.30  & 0.07\\
N2159  &   ...  &   11.38  & -0.34  & 0.16\\
N2160  &   ...  &    ...   &  ...   & 0.19\\
N2162  &   V    &   12.70  & -0.18  & 0.06\\
N2164  &   III  &   10.34  & -0.31  & 0.12\\
N2166  &   ...  &    ...   &  ...   & 0.22\\
N2172  &   ...  &   11.75  & -0.29  & 0.17\\
N2173  &   V-VI &   12.30  & -0.26: & 0.16:\\
N2177  &   ...  &    ...   &  ...   & 0.08\\
N2193  &   ...  &    ...   &  ...   & 0.33\\
N2209  &   III-I&V  13.15  & -0.18  & 0.04\\
N2210  &   VII  &   10.94  & -0.40  & 0.10\\
N2213  &   V-VI &   12.38  & -0.22  & 0.14:\\
N2214  &   II   &   10.93  & -0.35  & 0.22\\
N2231  &   V    &   13.20  & =3D0.21  & 0.10\\
N2249  &   ...  &   12,23  & =3D0.11  & 0.08\\
H11    &   VII  &   11.98  & -0.47  & 0.09\\
SL363  &   ...  &    ...   & -0.24  & 0.25\\
SL885  &   ...  &   14.3   &  ...   & 0.18\\

\enddata
\tablenotetext{*}{The cluster NGC 2109 is misidentified as NGC 2019 in Bhatia and MacGillivray (1989)} 
\end{deluxetable}


\begin{deluxetable}{llll}
\tablewidth{0pt}
\tablecaption{Flattening of clusters in the SMC measured at the half-light isophote} \tablehead{\colhead{Cluster} & \colhead{V} & \colhead{Q} & \colhead{$\epsilon$}}

\startdata

L1   &   13.32 & -0.37 & 0.09\\
L4   &    ...  &  ...  & 0.22\\
L6   &    ...  &  ...  & 0.25\\
L7   &    ...  &  ...  & 0.15\\
L8   &   12.05 & -0.36 & 0.10\\
L10  &   11.24 & -0.44 & 0.29\\
L12  &    ...  &  ...  & 0.15\\
L15  &   12.92 & -0.29 & 0.24\\
L16  &   12.70 & -0.38 & 0.28\\
L18  &    ...  &  ...  & 0.18\\
L20  &    ...  &  ...  & 0.15\\
L21  &    ...  &  ...  & 0.13\\
L23  &   14.30 &  ...  & 0.13\\
L27 &    13.66 &  ...  & 0.19\\
L28 &     ...  &  ...  & 0.14:\\
L29 &    12.03 & -0.41 & 0.15\\
L35 &    14.49 &  ...  & 0.22\\
L37 &    12.62 & -0.27 & 0.10\\
L47 &    12.26 & -0.19 & 0.18\\
L48 &    12.87 & -0.47 & 0.18\\
L53 &    12.53 & -0.18 & 0.15\\
L54 &     9.60 & -0.44 & 0.25\\
L57 &     ...  &  ...  & 0.03\\
L58 &    14.77 &  ...  & 0.25\\
L59 &    12.84 & -0.40 & 0.23\\
L67 &    12.78 & -0.42 & 0.16\\
L68 &    13.57 & -0.45 & 0.12\\
L72 &    10.75 & -0.40 & 0.25\\
L77 &     ...  &  ...  & 0.20\\
L80 &     ...  &  ...  & 0.20\\
L82 &    12.21 & -0.21 & 0.24\\
L83 &    11.42 & -0.39 & 0.08\\
L85 &    10.61 & -0.24 & 0.24\\

\enddata
\end{deluxetable}


\end{document}